\documentclass{article}

\begin{document}

\title{\bf \Large Classification of Spherically Symmetric Static Spacetimes according
to their Matter Collineations}

\author{M. Sharif \thanks{Present Address: Department of Mathematical Sciences,
University of Aberdeen, Kings College, Aberdeen AB24 3UE Scotland,
UK. $<$msharif@maths.abdn.ac.uk$>$} and Sehar Aziz
\\ Department of Mathematics, University of the Punjab,\\ Quaid-e-Azam Campus
Lahore-54590, PAKISTAN.}

\date{}

\maketitle

\begin{abstract}
{\it The spherically symmetric static spacetimes are classified
according to their matter collineations. These are studied when
the energy-momentum tensor is degenerate and also when it is
non-degenerate. We have found a case where the energy-momentum
tensor is degenerate but the group of matter collineations is
finite. For the non-degenerate case, we obtain either {\it four},
{\it five}, {\it six} or {\it ten} independent matter
collineations in which four are isometries and the rest are
proper. We conclude that the matter collineations coincide with
the Ricci collineations but the constraint equations are different
which on solving can provide physically interesting cosmological
solutions.}
\end{abstract}


\date{}

\newpage

\section{Introduction}

Einstein's field equations (EFEs) are given by
\begin{equation}
G_{ab}\equiv R_{ab}-\frac{1}{2}Rg_{ab}=\kappa T_{ab}, \quad (a,b=0,1,2,3),
\end{equation}
where $G_{ab}$ are the components of the Einstein tensor, $R_{ab}$
those of the Ricci and $T_{ab}$ of the matter (energy-momentum)
tensor. Also, $R = g^{ab} R_{ab}$ is the Ricci scalar, $\kappa$ is
the gravitational constant and, for simplicity, we take $\Lambda =
0$.

Let $(M,g)$ be a spacetime, i.e., $M$ is a four-dimensional,
Hausdorff, smooth manifold, and $g$ is smooth Lorentz metric of
signature (+ - - -) defined on $M$. The manifold $M$ and the
metric $g$ are assumed smooth ($C^{\infty}$). We shall use the usual
component notation in local charts, and a covariant derivative with
respect to the symmetric connection $\Gamma$ associated with the metric
$g$ will be denoted by a semicolon and a partial derivative by a comma.

In general relativity (GR) theory, the Einstein tensor $G_{ab}$ plays
a significant role, since it relates the geometry of spacetime to
its source. The GR theory, however, does not prescribe the various
forms of matter, and takes over the energy-momentum tensor
$T_{ab}$ from other branches of physics.

Collineations are geometrical symmetries which are defined by a
relation of the form
\begin{equation}
\pounds_{\xi}\Phi=\Lambda,
\end{equation}
where $\pounds$ is the Lie derivative operator, $\xi^a$ is the
symmetry or collineation vector, $\Phi$ is any of the quantities
$g_{ab},\Gamma^a_{bc},R_{ab},R^a_{bcd}$ and geometric objects
constructed by them and $\Lambda$ is a tensor with the same index
symmetries as $\Phi$. One can find all the well-known
collineations by requiring the particular forms of the quantities
$\Phi$ and $\Lambda$. For example if we take $\Phi_{ab}=g_{ab}$
and $\Lambda_{ab}=2\psi g_{ab}$, where $\psi(x^a)$ is a scalar
function, this defines a Conformal Killing vector (CKV) and it
specializes to a Special Conformal Killing vector (SCKAV) when
$\psi_{;ab}=0$, to a Homothetic vector field when $\psi=constant$
and to a Killing vector (KV) when $\psi=0$. If we take
$\Phi_{ab}=R_{ab}$ and $\Lambda_{ab}=2\psi R_{ab}$ the symmetry
vector $\xi^a$ is called a Ricci inheritance collineation (RIC)
and reduces to a Ricci collineation (RC) for $\Lambda_{ab}=0$.
When $\Phi_{ab}=T_{ab}$ and $\Lambda_{ab}=2\psi T_{ab}$, where
$T_{ab}$ is the energy-momentum tensor, the vector $\xi^a$ is
called a Matter inheritance collineation (MIC) and it reduces to a
Matter collineation (MC) for $\Lambda_{ab}=0$. In the case of
CKVs, the function $\psi$ is called the conformal factor and in
the case of inheriting collineations the inheriting factor.

Collineations can be proper (non-trivial) or improper (trivial).
In this paper, we will define a proper MC to be an MC which is not
a KV, or a HV. The MC equation can be written as
\begin{equation}
\pounds_{\xi} T_{ab} = 0 \quad \Leftrightarrow \quad \pounds_{\xi}
G_{ab} = 0,
\end{equation}
or in component form
\begin{equation}
T_{ab,c} \xi^c + T_{ac} \xi^c_{,b} + T_{cb} \xi^c_{,a} = 0.
\end{equation}

Collineations other than motions (KVs) can be considered as non-Noetherian
symmetries and can be associated with constants of motion and, up to the
level of CKVs, they can be used to simplify the metric [1]. For example,
Affine vectors (AVs) are related to conserved quantities [2], RCs, are
related to the conservation of particle number in Friedmann Robertson-Walker
spacetimes [3]. and the existence of Curvature collineations (CCs) implies
conservation laws for null electromagnetic fields [4]. The set of
collineations of a spacetime can be related with an inclusion relation
leading to a tree like inclusion diagram [4] which shows their relative hierarchy.
A collineation of a given type is proper if it does not belong to any of
the subtypes in this diagram. In order to relate a collineation to a particular
conservation law and its associated constant(s) of motion, the properness of
the collineation must first be assured.

The motivation for studying MCs can be discussed as follows. When
we find exact solutions to the Einstein's field equations, one of
the simplifications we use is the assumption of certain symmetries
of the spacetime metric. These symmetry assumptions are expressed
in terms of isometries expressed by the spacetimes, also called
Killing vectors which give rise to conservation laws [1,5].
Symmetries of the energy-momentum tensor provide conservation laws
on matter fields. These symmetries are called matter
collineations. These enable us to know how the physical fields,
occupying in certain region of spacetimes, reflect the symmetries
of the metric [6]. In other words, given the metric tensor of a
spacetime, one can find symmetry for the physical fields
describing the material content of that spacetime. There is also a
purely mathematical interest of studying the symmetry properties
of a given geometrical object, namely the Einstein tensor, which
arises quite naturally in the theory of GR. Since it is related,
via the Einstein field equations, to the material content of the
spacetime, it has an important role in this theory.

Recently, there is a growing interest in the study of MCs [7-13].
Carot, et al [8] has discussed MCs from the
point of view of the Lie algebra of vector fields generating them
and, in particular, he discussed spacetimes with a degenerate
$T_{ab}$. Hall, et al [9], in the discussion of RC and MC, have
argued that the symmetries of the energy-momentum tensor may also
provide some extra understanding of the the subject which has not
been provided by Killing vectors, Ricci and Curvature
collineations.

In this paper, we study the problem of calculating MCs for static
spherically symmetric spacetimes for both degenerate and
non-degenerate energy-momentum tensors and establish the relation
between KVs, RCs and MCs. The breakdown of the paper follows. In
the next section we write down MC equations for static spherically
symmetric spacetimes. In section three, we shall solve these MC
equations when the energy-momentum tensor is degenerate and in the
next section MC equations are solved for the non-degenerate
energy-momentum tensor. Finally, a summary of the results obtained
will be presented.

\section{Matter Collineation Equations}

In this section, we write down the MC equations for spherically
symmetric static spacetimes. The most general spherically
symmetric metric is given as
\begin{equation}
ds^2 = e^{\nu(t,r)}dt^2-e^{\lambda(t,r)}d r^2-e^{\mu(t,r)}d\Omega^2,
\end{equation}
where $d\Omega^2=d\theta^2+\sin^2 \theta d\theta^2$. Since we are
dealing with static spherically symmetric spacetimes, Eq.(5)
reduces to
\begin{equation}
ds^2 = e^{\nu(r)}dt^2-e^{\lambda(r)}d r^2-e^{\mu(r)}d\Omega^2,
\end{equation}
We can write MC Eqs.(4) in the expanded form as follows
\begin{equation}
T_{0,1} \xi^1 + 2 T_0 \xi^0_{,0} = 0,
\end{equation}
\begin{equation}
T_0 \xi^0_{,1} + T_1 \xi^1_{,0} = 0,
\end{equation}
\begin{equation}
T_0 \xi^0_{,2} + T_2 \xi^2_{,0} = 0,
\end{equation}
\begin{equation}
T_0 \xi^0_{,3} + \sin^2 \theta T_2 \xi^3_{,0} = 0,
\end{equation}
\begin{equation}
T_{1,1} \xi^1 + 2 T_1 \xi^1_{,1} = 0,
\end{equation}
\begin{equation}
T_1 \xi^1_{,2} + T_2 \xi^2_{,1} = 0,
\end{equation}
\begin{equation}
T_1 \xi^1_{,3} +\sin^2 \theta T_2 \xi^3_{,1} =0,
\end{equation}
\begin{equation}
T_{2,1} \xi^1 + 2 T_2 \xi^2_{,2} =0,
\end{equation}
\begin{equation}
\xi^2_{,3} + \sin^2 \theta \xi^3_{,2} =0,
\end{equation}
\begin{equation}
T_{2,1} \xi^1 + 2 T_2 \cot \theta \xi^2 + 2 T_2\xi^3_{,3} = 0,
\end{equation}
where $T_{3}=\sin^2\theta T_{2}$. It is to be noticed that we are
using the notation $T_{aa}=T_a$. We solve these equations for the
degenerate as well as the non-degenerate case. The nature of the
solution of these equations changes when one (or more) $ T_{a} $
is zero. The nature changes even if $ T_{a} \neq 0$ but $ T_{a,1}
=0$.

\section{Matter Collineations in the Degenerate Case}

In this section we solve MC equations (7)-(16) when at least one
of $T_a=0$. First, we consider the trivial case, where $T_a=0$. In
this case, Eqs.(7)-(16) are identically satisfied and thus every
vector field is an MC.

The other possibilities can be classified in three main cases:
\bigskip

({\it 1}) when only one of the $T_a \neq 0$,

({\it 2}) when exactly two of the $T_a \neq 0$,

({\it 3}) when exactly three of the $T_a \neq 0$.
\bigskip
\par \noindent
\par \noindent
{\bf Case ({\it 1})}: In this case, there could be only two
possibilities:

$({\it 1a}):T_0 \neq 0,\quad T_{i} = 0 \quad (i = 1,2,3),~~ ({\it
1b}): T_1 \neq 0, \quad T_{j} = 0 \quad (j = 0,2,3).$
\par \noindent
\par \noindent
The case ({\it 1a}) is trivial and we get either ({\it i})
$T_0=constant\neq 0$ or ({\it ii}) $T_0\neq constant$. For the
first possibility, we have
\begin{equation}
\xi = c_0\partial_t+\xi^i(x^a)\partial_i,
\end{equation}
where $c_0$ is a constant. For the second possibility, we obtain
\begin{equation}
\xi = f(t)\partial_t-{\dot{f(t)}\over[\ln\sqrt{T_{0}}]'}\partial_r
+\xi^{\ell}(x^a)\partial_\ell,~~\ell=2,3.
\end{equation}
In the case ({\it 1b}), MC Eqs.(7), (9), (10) and (14)-(16) are identically
satisfied while Eqs.(8), (12), (13) yield that $\xi^1=\xi^1(r)$ only. Using
Eq.(11), we have the following solution
\begin{equation}
\xi = \frac{c_0}{\sqrt{T_1}}\partial_r+\xi^j(x^a)\partial_j.
\end{equation}
We have seen that in all subcases of the case (1), we obtain
infinite number of MCs.
\par \noindent
\par \noindent
{\bf Case ({\it 2})}: This case implies the following two
possibilities:

$(2a):  T_k = 0, \quad T_{\ell} \neq 0,~~ (2b): T_k \neq 0, \quad
T_{\ell} = 0,$
\par \noindent
\par \noindent
where $k = 0, 1$ and $\ell = 2,3$ which are valid through this
case.
\par \noindent
\par \noindent
The case ({\it 2a}) explores further two possibilities i.e. ({\it
i}) $T_2=constant\neq 0$ and ({\it ii}) $T_2\neq constant$. In the
first option, we have the following solution of the MC equations
\begin{equation}
\xi^k = \xi^k(t,r,\theta,\phi),~~ \xi^2 = c_1 \cos \phi + c_2 \sin
\phi,~~ \xi^3 = - \cot \theta (c_1 \sin \phi - c_2 \cos \phi) +
c_0,
\end{equation}
where $c_0$, $c_1$, $c_2$ are constants. Thus we can write
\begin{equation}
\xi = \xi^k(x^a)\partial_k+c_1 (\cos \phi\partial_\theta- \cot
\theta\sin\phi\partial_\phi) + c_2(\sin\phi\partial_\theta+\cot
\theta\cos\phi\partial_\phi)+c_0\partial_\phi.
\end{equation}
We see that this contains the KVs associated with the usual
spherical symmetry given in Appendix B.
\par \noindent
\par \noindent
In the case ({\it 2aii}), MCs turn out to be
\begin{equation}
\xi = \xi^0(x^a)\partial_t+c_1 (\cos \phi\partial_\theta- \cot
\theta\sin\phi\partial_\phi) + c_2(\sin\phi\partial_\theta+\cot
\theta\cos\phi\partial_\phi)+c_0\partial_\phi.
\end{equation}
We again have the usual three KVs in addition to arbitrary MCs in
the $t$ direction.
\par \noindent
\par \noindent
For the case ({\it 2b}), Eqs.(14)-(16) are identically satisfied
and from the remaining Eqs.(7)-(13), we have
\begin{equation}
{\ddot{A} \over A} = {T_{0} \over T_1} \left( {T_{0,1} \over
2 T_{0} \sqrt{T_{1}} } \right)' = \alpha,
\end{equation}
where $ \alpha $ is a separation constant. From Eq.(22), three
possibilities arise

$({\it i})~\alpha > 0,~~({\it ii})~ \alpha = 0,~({\it iii})~ \alpha < 0.$
\par \noindent
\par \noindent
Solving MC equations for the case ({\it 2bi}), we have the following set of MCs
\begin{equation}
\xi^0 = - {T_{0,1} \over 2 T_{0} \sqrt{T_{1}} \sqrt{\alpha} } (c_1
\sin h \sqrt{\alpha} t + c_2 \cos h \sqrt{\alpha} t) + c_0,
\end{equation}
\begin{equation}
\xi^1 = {c_1 \cos h \sqrt{\alpha} t + c_2 \sin h \sqrt{\alpha} t
\over \sqrt{T_{1}} },~~\xi^{\ell} =\xi^{\ell}(t,r,\theta,\phi).
\end{equation}
In the case of ({\it 2bii}), we obtain the following solution
\begin{equation}
\xi^0 = - \beta (c_1 {t^2 \over 2} + c_2 t) - c_1 \int
{\sqrt{T_{1}} \over T_{0} } dr + c_0,
\end{equation}
\begin{equation}
\xi^1 ={c_1 t + c_2 \over \sqrt{T_{1}} },~~\xi^{\ell}
=\xi^{\ell}(t,r,\theta,\phi),
\end{equation}
where $ \beta $ is an integration constant which can be zero or
non-zero and is given by
\begin{equation}
{T_{0,1} \over 2 T_{0} \sqrt{T_{1}} } = \beta,
\end{equation}
For the case of ({\it 2biii}), solution of MC equations becomes
\begin{equation}
\xi^0 = \sqrt{p}(c_1 \sin \sqrt{p} t - c_2 \cos \sqrt{p} t) \int
{\sqrt{T_{1}} \over T_{0} } d r - c_0,
\end{equation}
\begin{equation}
\xi^1 = {c_1 \cos \sqrt{p} t + c_2 \sin \sqrt{p} t \over
\sqrt{T_{1}} },~~\xi^{\ell} =\xi^{\ell}(t,r,\theta,\phi),
\end{equation}
where $\alpha=-p$ and $p>0$. When $T_0$ and $T_1$ do not satisfy
Eq.(22), we have the following solution
\begin{equation}
\xi=c_0\partial_t+\xi^{\ell}(x^a)\partial_{\ell}.
\end{equation}
Again we see that all the possibilities of the case (2) give
infinite-dimensional MCs.
\par \noindent
\par \noindent
{\bf Case ({\it 3})}: In this case, we have the following two
possibilities:

$({\it 3a}): T_0 = 0, \quad T_{i} \neq  0,~~ ({\it 3b}): T_1 = 0,
\quad T_{j} \neq 0.$
\par \noindent
\par \noindent
For the case ({\it 3a}), Eq.(7) is identically satisfied
and Eqs.(8)-(10) respectively give
$\xi^i = \xi^i (r,\theta,\phi),~(i=1,2,3)$. From the remaining equations,
we have the following constraint
\begin{equation}
{A_{,22} \over A} = {T_{2} \over \sqrt{T_{1}} } \left( { T_{2,1}
\over 2 T_{2} \sqrt{T_{1}} } \right)' =\alpha,
\end{equation}
where $ \alpha $ is a separation constant. This implies that we have
three different possibilities:

({\it i})~~$ \alpha > 0 $,\quad({\it ii})~~$ \alpha = 0 $,\quad
({\it iii})~~$ \alpha < 0$.
\par \noindent
\par \noindent
The case ({\it 3ai}) gives the same MCs as for the case (2aii).
\par \noindent
\par \noindent
For the case ({\it 3aii}), we have two possibilities depending
upon the value of constraint
$\beta=\frac{T_{2,1}}{2T_2\sqrt{T_1}}$. When $\beta=0$, we have
the following MCs
\begin{eqnarray}
\xi = \xi^0(x^a)\partial_t+c_0\frac{1}{\sqrt{T_1(r)}}\partial_r+
c_1(\cos \phi\partial_\theta- \cot \theta\sin\phi\partial_\phi) +
c_2(\sin\phi\partial_\theta\nonumber\\+\cot
\theta\cos\phi\partial_\phi)+c_3\partial_\phi.
\end{eqnarray}
This again yields infinite-dimensional MCs in addition to the
usual spherical symmetry KVs. For $\beta\neq 0$, MCs turn out to
be of the case (2aii).
\par \noindent
\par \noindent
The case ({\it 3aiii}) gives the following MCs
\begin{eqnarray}
\xi=\xi^0\partial_t+c_0(\frac{\cos\theta}{\sqrt{T_1(r)}}\partial_r
-\frac{T'_2(r)}{2T_2(r)\sqrt{T_2(r)}}\sin\theta\partial_\theta)\nonumber\\
+c_1[\frac{\sin\theta\cos\phi}{\sqrt{T_1(r)}}\partial_r
+\frac{T'_2(r)}{2T_2(r)\sqrt{T_2(r)}}(\cos\theta\cos\phi\partial_\theta
+\sin\phi\partial_\phi)]\nonumber\\
+c_2[\frac{\cos\theta\sin\phi}{\sqrt{T_1(r)}}\partial_r
+\frac{T'_2(r)}{2T_2(r)\sqrt{T_2(r)}}(\cos\theta\sin\phi\partial_\theta
-\cos\phi\partial_\phi)]\nonumber\\
+c_3\frac{T'_2(r)}{2T_2(r)\sqrt{T_2(r)}}(\cos\phi\partial_\theta
-\cot\theta\sin\phi\partial_\phi)\nonumber\\
+c_4\frac{T'_2(r)}{2T_2(r)\sqrt{T_2(r)}}(\sin\phi\partial_\theta
+\cot\theta\cos\phi\partial_\phi)\nonumber\\
+c_5\frac{T'_2(r)}{2T_2(r)\sqrt{T_2(r)}}\partial_\phi.
\end{eqnarray}
This again provides infinite-dimensional MCs in addition to the
usual KVs.
\par \noindent
\par \noindent
The case ({\it 3b}) deals with the constraints $ T_{0} \neq
0$,\quad $T_{1} = 0$ and $T_{2} \neq 0$. In addition to these, we
can have the following constraints

({\it i}) $T_{j, 1} \neq 0,$~~
({\it ii}) $ T_{0, 1} = 0,~  T_{2,1} \neq 0$,~~

({\it iii}) $ T_{0, 1} \neq 0,~ T_{2,1} = 0$,~~
({\it iv}) $ T_{0, 1} = 0,~~ T_{2,1} = 0.$
\par \noindent
\par \noindent
In case ({\it 3bi}), from MC Eqs.(7)-(16), we obtain the following MCs
\begin{equation}
\xi =
c_0\partial_t+c_1(\cos\phi\partial_\theta-\cot\theta\sin\phi\partial_\phi)
+c_2(\sin\phi\partial_\theta+\cot\theta\cos\phi\partial_\phi)
+c_3\partial_\phi
\end{equation}
which gives four independent MCs. This case is worth mentioning as
we have found finite number of MCs even for the degenerate
energy-momentum tensor.
\par \noindent
\par \noindent
In case ({\it 3bii}), Eq.(11) becomes identity while Eqs.(7) and
(8) give $\xi^0=\xi^0(\theta,\phi)$. Also, Eqs.(12) and (13)
respectively show that $\xi^2$ and $\xi^3$ are functions of $t,
\theta,\phi$. From the remaining MC equations, we have the
following solution
\begin{eqnarray}
\xi =
c_0\partial_t+c_1(\cos\phi\partial_\theta-\cot\theta\sin\phi\partial_\phi)
+c_2(\sin\phi\partial_\theta+\cot\theta\cos\phi\partial_\phi)
+c_3\partial_\phi\nonumber\\
-\left[\frac{1}{\ln{\sqrt{T_2}}}\frac{\partial}{\partial\theta}
\{[f_+(z)+f_-(\bar{z})]\sin\theta\}\right]\partial_r
+[f_+(z)+f_-(\bar{z})]\sin\theta\partial_{\theta}\nonumber\\
-[f_+(z)+f_-(\bar{z})]\partial_\phi,
\end{eqnarray}
where $z=\ln\mid\csc\theta-\cot\theta\mid+\iota\phi$ and $f_+(z)$
and $f_-(\bar{z})$ are arbitrary functions of $z$ and $\bar{z}$
(the complex conjugate of $z$) such that their sum is real and
difference is imaginary.
\par \noindent
\par \noindent
For the case ({\it 3biii}), one obtains the following MCs
\begin{eqnarray}
\xi = f(t)\partial_t+{- \dot{f} (t) \over [\ln \sqrt{T_{0}} ]'
}\partial_r+c_1(\cos\phi\partial_\theta-\cot\theta\sin\phi\partial_\phi)
+c_2(\sin\phi\partial_\theta\nonumber\\+\cot\theta\cos\phi\partial_\phi)
+c_3\partial_\phi
\end{eqnarray}
which gives an infinite number of MCs.
\par \noindent
\par \noindent
The case ({\it 3biv}) yields the following MC vectors
\begin{equation}
\xi = c_0\partial_t+f
(x^a)\partial_r+c_1(\cos\phi\partial_\theta-\cot\theta\sin\phi\partial_\phi)
+c_2(\sin\phi\partial_\theta+\cot\theta\cos\phi\partial_\phi)
+c_3\partial_\phi.
\end{equation}
Thus one has an infinite number of independent MCs in all subcases
of of the case (3) except for (3bi) which has four independent
MCs.

\section{Matter Collineations in the Non-degenerate Case}

In this section we shall solve MC equations when the energy-momentum tensor
is non-degenerate, i.e, $ T_{a} \neq 0$ since $det(T_a)\neq 0$.
\par \noindent
\par \noindent
If we solve MC Eqs.(7)-(16), after some tedious algebra, we arrive
at the following solution
\begin{eqnarray}
\xi^0 & = &- {T_2\sin \theta \over T_0 } [\dot{A_1}\sin \phi -
\dot{A_2}\cos \phi] + {T_2\over T_1}\dot{A_3}\cos \theta
+ A_4, \\
\xi^1 & = & -{T_2\sin \theta \over T_1 } [A_1' \sin \phi -
A'_2 \cos \phi] + {T_2 \over T_1}A'_3\cos \theta
+ A_5, \\
\xi^2 & = & \cos \theta [A_1 \sin \phi -
A_2 \cos \phi] + A_3\sin \theta+ c_1 \sin \phi - c_2 \cos \phi, \\
\xi^3 & = & {\rm cosec} \theta [A_1 \cos \phi +
A_2 \sin \phi] + [c_1 \cos \phi + c_2 \sin \phi]\cot \theta+ c_0 ,
\end{eqnarray}
where $c_0, c_1$ and $c_2$ are arbitrary constants and
$A_\mu=A_\mu(t,r), \mu=1,2,3,4,5$. These $\xi^a$ are satisfied
subject to the following differential constraints on $A_\mu$
\begin{eqnarray}
2 T_1 \ddot{A}_i+T_{0,1}A'_i=0,~~i = 1,2,3,\\
2T_0 \dot{A_4}+T_{0,1}A_5 =  0,\\
2T_2 \dot{A'_i} + T_{0} \left({T_{2} \over T_{0} } \right)' \dot{A_i}
= 0,\\
T_{0} A'_4 + T_{11} \dot{A_5} = 0,\\
\left\{T_{1,1} {T_2 \over T_1} + 2 T_1 \left({ T_2
\over T_1} \right)' \right\} A'_i + 2 T_2 A_i^{''} = 0,\\
T_{1,1} A_5 + 2 T_1 A'_5 = 0,\\
T_{2,1} A'_i + 2 T_1 A_i = 0,~~c_0 = 0,\\
T_{2,1} A_5 = 0.
\end{eqnarray}
Now the problem of working out MCs for all possibilities of $ A_i,
A_4, A_5 $ is reduced to solving the set of Eqs.(39)-(42) subject
to the above constraints. We start the classification of MCs by
considering the constraint Eq.(50). This can be satisfied for
three different possible cases.
\begin{eqnarray*}
{\it (1)}~~&& T_{2,1} = 0,~~~A_5 \neq 0,\\
{\it (2)}~~& & T_{2,1} \neq 0,~~~A_5 = 0,\\
{\it (3)}~~& & T_{2,1} = 0,~~~A_5 = 0.
\end{eqnarray*}
{\bf Case {\it (1)}}: In this case, all the constraints remain
unchanged except (43), (47) and (49). Thus we have
\begin{eqnarray}
\dot{A'_i} - {1 \over 2} {T_{0,1} \over T_{0} } \dot{A_i} = 0,\\
A''_i - {T_{1,1} \over 2 T_{1} } A'_i = 0, \\
T_{1} A_i = 0.
\end{eqnarray}
The last equation is satisfied only if $ A_i = 0. $ As a result,
all the differential constraints involving $ A_i $ and its
derivatives disappear identically and we are left with Eqs.(44),
(46) and (48) only. Now integrating constraint Eq.(48) w.r.t. $ r
$ and replacing the value of $ A_5 $ in constraint Eq.(44), we
have
$$ T_{0,1} {A(t) \over \sqrt{T_{1}} } + 2T_{0} \dot{A_4} = 0,$$
where $ A(t) $ is an integration function. This can be satisfied
for the following two possibilities:
\par \noindent
\par \noindent
\begin{eqnarray*}
(a)~~T_{0,1} = 0,~~~\dot{A_4} = 0,~~~(b)~~T_{0,1} \neq
0,~~~\dot{A_4}\neq 0.
\end{eqnarray*}
For the case (1a), after some algebra, we arrive at the following
set
\begin{eqnarray}
\xi = c_0\partial_t+c_4[\{{-1 \over a} \int
\sqrt{T_{1}}dr\partial_t
+{t\over \sqrt{T_{1}} }\}\partial_r]+c_5 {1\over \sqrt{T_{1}}}
\partial_r\nonumber\\
+c_1(\cos\phi\partial_\theta-\cot\theta\sin\phi\partial_\phi)
+c_2(\sin\phi\partial_\theta+\cot\theta\cos\phi\partial_\phi)
+c_3\partial_\phi,
\end{eqnarray}
where $a,c_0,c_1,c_2,c_3,c_4,c_5$ are arbitrary constants. This
shows that we have six MCs.
\par \noindent
\par \noindent
In the case (1b), we have $ \dot{A_4} \neq 0 $ and $ T_{0,1} = 0
$. Solving Eqs.(44) and (46) and rearranging terms, we get
\begin{equation}
{\ddot{A} \over A} = {1 \over 2} \left[ { T_{0,1} \over T_{0}
\sqrt{T_{1}} } \right]' {T_{0} \over \sqrt{T_{1}} } = \alpha,
\end{equation}
where $ \alpha $ is a separation constant and this gives the
following  three possible cases:
$$(i)~~\alpha < 0,~~~ (ii)~~\alpha = 0,~~(iii)~~
\alpha> 0. $$
\par \noindent
\par \noindent
First we take $ \alpha < 0 $ and assume that $ \alpha = - p $,
where $ p$ is positive, then Eq.(55) yields
$$A = A_{11} \cos \sqrt{p} t + A_{12} \sin \sqrt{p} t $$
and also
\begin{equation}
{1 \over 2} \left[ {T_{0,1} \over T_0 \sqrt{T_1} } \right]' = - {p
\sqrt{T_1} \over T_0 }.
\end{equation}
After some algebra, we can write the values of $A_4$ and $A_5$
given as
$$ A_4 = \sqrt{p} \int {\sqrt{T_{1}} \over T_{0} } d r (A_{11} \sin
\sqrt{p} t - A_{12} \cos \sqrt{p} t) + c_0, $$
$$ A_5 = { A_{11} \cos \sqrt{p} t + A_{12} \sin \sqrt{p} t \over
\sqrt{T_{1}} }. $$ Substituting these values in Eqs.(39)-(42) and
using the constraints $ A_{11} = 0 = A_{12} $, we obtain four
independent MCs which are exactly similar to the case (3bi).
\par \noindent
\par \noindent
The subcase (1bii) gives
$$ A(t) = c_3 t + c_4 $$
and
\begin{equation}
{T_{0,1} \over T_{0} \sqrt{T_{1}} } = \beta,
\end{equation}
where $\beta$ is an integration constant which yields the
following two possibilities
$$ (*)~~ \beta \neq 0,~~~~~~(**)~~ \beta = 0. $$
\par \noindent
\par \noindent
The first possibility implies that $$ T_{0} = \beta_0 e^{\beta
\int \sqrt{T_{1}} dr},$$ where $ \beta_0 $ is an integration
constant. Now we solve Eqs.(48) and (50) by using this constraint,
we can get the following set
\begin{eqnarray}
\xi^0 = - {c_3 \over \beta_0} \int {\sqrt{T_{1}} \over e^{\beta
\int \sqrt{T_{1}} d r} }d r + c_0,~~
\xi^1 = {c_3 t + c_4 \over \sqrt{T_{1}}}, \nonumber\\
\xi^2 = c_1 \sin \phi - c_2 \cos \phi,~~ \xi^3 = c_1 \cos \phi +
c_2 \sin \phi.
\end{eqnarray}
This gives five independent MCs.
\par \noindent
\par \noindent
For the case (1bii**), $T_0=constant$. Using this fact Eq.(48)
yields $A_4=g(r)$. Thus we have the solution
\begin{eqnarray}
\xi^0 = - \frac{c_3}{b}\int{\sqrt{T_1}dr} + c_0,~~
\xi^1 = {c_3 t + c_4 \over \sqrt{T_1}}, \nonumber\\
\xi^2 = c_1 \sin \phi - c_2 \cos \phi,~~ \xi^3 = c_1 \cos \phi +
c_2 \sin \phi.
\end{eqnarray}
We again have five MCs.
\par \noindent
\par \noindent
The case (1biii) will give the similar results as in the case
(1bi).
\par \noindent
\par \noindent
{\bf Case (\it{2})}: In this case, Eqs.(44) and (46) show that
$A_4 $ is a pure constant. Integration of constraint Eq.(49)
w.r.t. $ r $ gives
\begin{equation}
A_i = c_{i1} (t)e^{-2 \int {T_{1} \over T_{2,1} } d r},
\end{equation}
where $ c_{i1} (t) $ is integration function. Substituting this in
Eq.(43), we have after some simplifications
\begin{equation}
\frac{\ddot{c}_{i1}(t)}{c_{i1}} = {T_{0,1} \over T_{2,1} } =
\alpha,
\end{equation}
where $ \alpha $ is a separation constant. From here we have three
possibilities

$$ (a) ~~\alpha > 0,~~~(b)~~ \alpha =0,~~~~(c)~~ \alpha < 0. $$
\par \noindent
\par \noindent
In the case (2a), Eq.(61) gives
$$c_{i1}=a_{i1}\cosh\sqrt{\alpha}t+a_{i2}\sinh\sqrt{\alpha}t.$$
Substituting this value in Eq.(60), we get
$$A_i=(a_{i1}\cosh\sqrt{\alpha}t+a_{i2}\sinh\sqrt{\alpha}t)
e^{-2 \int {T_{1} \over T_{2,1} } d r}.$$ Replacing this value in
Eqs.(39)-(42) and then substituting the resulting values of
$\xi^a$ in MC Eq.(8), there are two possibilities either
$a_{1i}=a_{2i}=a_{3i}=0$ or
$$-4T_0T_1T_2+T_0T_{2,1}^2-T_0T_2T_{0,1}^2=0.$$ For the first
possibility we obtain the result as for the case (1bi). For the
second possibility subject to the constraint
$$T_2T_{1,1}T_{2,1}-4T_1^2T_2T_{2,1}+2T_1T_{2,1}^2-2T_1T_2T_{2,11}=0,$$
we have the following results
\begin{eqnarray}
\xi^0 = {T_{2} \over T_0}e^{-2 \int {T_{1} \over T_{2,1} } d
r}\sqrt{\alpha}[-\sin\theta
\{(a_{11}\sinh\sqrt{\alpha}t+a_{12}\cosh\sqrt{\alpha}t)\sin
\phi\nonumber\\-(a_{21}\sinh\sqrt{\alpha}t
+a_{22}\cosh\sqrt{\alpha}t)\cos\phi\}
+\cos\theta(a_{31}\sinh\sqrt{\alpha}t\nonumber\\+a_{32}\cosh\sqrt{\alpha}t)]+ c_0,
\nonumber\\
\xi^1 ={-2T_{2} \over T_{2,1}}e^{-2 \int {T_{1} \over T_{2,1} }dr}
[-\sin\theta
\{(a_{11}\cosh\sqrt{\alpha}t+a_{12}\sinh\sqrt{\alpha}t)\sin
\phi\nonumber\\-(a_{21}\cosh\sqrt{\alpha}t+a_{22}\sinh\sqrt{\alpha}t)\cos
\phi\}
+\cos\theta(a_{31}\cosh\sqrt{\alpha}t\nonumber\\+a_{32}\sinh\sqrt{\alpha}t)]+ c_0,
\nonumber\\
\xi^2 =(e^{-2 \int {T_{1} \over T_{2,1} } d r})
[\cos\theta\{(a_{11}\cosh\sqrt{\alpha}t+a_{12}\sinh\sqrt{\alpha}t)\sin
\phi\nonumber\\-(a_{21}\cosh\sqrt{\alpha}t+a_{22}\sinh\sqrt{\alpha}t)\cos
\phi\}+\sin\theta(a_{31}\cosh\sqrt{\alpha}t\nonumber\\+a_{32}\sinh\sqrt{\alpha}t)]
+ (c_1 \sin\phi - c_2 \cos \phi),\nonumber\\
\xi^3 = \cot \theta [c_1 \cos \phi + c_2 \sin \phi] + c_3 +
\csc\theta( e^{-2 \int {T_{1} \over T_{2,1} } d r})[
(a_{11}\cosh\sqrt{\alpha}t\nonumber\\+a_{12}\sinh\sqrt{\alpha}t)\cos\phi
+(a_{21}\cosh\sqrt{\alpha}t+a_{22}\sinh\sqrt{\alpha}t)\sin \phi ].
\end{eqnarray}
For the case (2b), Eq.(61) gives $T_0=constant$ and
$$ c_{i1} = a_{i1} t + a_{i2}, $$
Using the value of $c_{i1}$ in Eq.(60), we obtain
\begin{equation}
A_i = (a_{i1} t + a_{i2})e^{-2 \int {T_{1} \over T_{2,1} } d r}.
\end{equation}
Plugging these values in Eqs.(39)-(42) and re-labelling
$$ a_{11} = c_{4},~~~a_{21} = c_5,~~~a_{31} = c_6,~~~~a_{12} =
c_7,~~~~ a_{22} = c_8~~~~a_{32} = c_9,$$ we obtain the following
MCs
\begin{eqnarray}
\xi^0 = -\frac{T_2}{a}\left[\left\{c_4\sin\phi-c_5\cos\phi\right\}
\sin \theta-c_6\cos\theta\right]e^{-2 \int
{T_{1} \over T_{2,1} } d r } + c_0,\nonumber\\
\xi^1 = 2\frac{T_2}{T_{2,1}}[\{(c_4 t + c_7) \sin \phi - (c_5 t +
c_8) \cos \phi \} \sin \theta-(c_6 t \nonumber\\+ c_9) \cos
\theta]e^{-2 \int {T_{1} \over T_{2,1} } d r
},\nonumber\\
\xi^2 = [\{ (c_4 t + c_7) \sin \phi - (c_5 t + c_8) \cos \phi\}
\cos \theta + (c_6 t + c_9) \sin \theta]e^{-2 \int {T_{1} \over
T_{2,1} } d r} \nonumber\\+ (c_1 \sin \phi - c_2
\cos \phi),\nonumber\\
\xi^3 = \cot \theta [c_1 \cos \phi + c_2 \sin \phi] + c_3 +
\csc\theta[ (c_4 t + c_7) \cos \phi + (c_5 t + c_8) \sin \phi ].
\end{eqnarray}
In the case (2c), Eq.(61) gives
$$c_{i1}=a_{i1}\cos\sqrt{\alpha}t+a_{i2}\sin\sqrt{\alpha}t.$$
Substituting this value in Eq.(60), we get
$$A_i=(a_{i1}\cos\sqrt{\alpha}t+a_{i2}\sin\sqrt{\alpha}t)
e^{-2 \int {T_{1} \over T_{2,1} } d r}.$$ Further this case
proceeds in the same way as the case (2a) and consequently the
equivalent results.
\par \noindent
\par \noindent
{\bf Case (\it {3})}: In this case, we have $T_2=constant$ and
$A_5= 0$. This can be solved trivially and gives similar results
as in the case (1bi).

\section{Discussions and Conclusions}

In the classification of spherically symmetric static spacetimes
according to the nature of the energy-momentum tensor, we find
that when the energy-momentum tensor is degenerate, Sec. (3), then
there are many cases where the MCs are infinite-dimensional. It is
very interesting to note that we have found a case ({\it 3bi})
where the energy-momentum tensor is degenerate but the group of
MCs is finite dimensional, i.e., there are four independent MCs.
In the cases (1)-(3), we summarize some results in the following:
\begin{description}
\item \qquad {\bf 1}. In this case, i.e., the rank of $T_{ab}$
being 1, it is found that all the possibilities yield
infinite-dimensional MCs.

\item \qquad {\bf 2}. In all subcases of this case, the rank of
$T_{ab}$ being 2. In subcase (2b), solving the equations $T_2 = 0$
and $T_3 = 0$, we obtain the Bertotti-Robinson I metric [13] given
by
\begin{eqnarray}
ds^2 &=& (B+r)^2dt^2 - dr^2 - a^2d\Omega^2,
\end{eqnarray}
where $B$ and $a$ are constants. This metric has six KVs but MCs are
infinite dimensional.

\item \qquad {\bf 3}. For this case, the rank of $T_{ab}$ is 3.
The point worth mentioning in this case is the case (3bi) in which
we have finite dimensionality of the group of MCs even if the
energy-momentum tensor is degenerate. We obtain {\it four} MCs
including the three KVs of spherical symmetric spacetimes.
\end{description}

Furthermore, we have dealt with the case when the energy-momentum
tensor is non-degenerate (Sec.4). There are three main categories
of the non-degenerate case which can be summarized as follows:
\begin{description}
\item \qquad {\bf 1}. This case yields two possibilities. In the
first possibility, we obtain six independent MCs in which four are
the KVs and two are the proper MCs. The second possibility further
has a division of three cases. The first case turns out similar to
the case (3bi) of degenerate case which gives four MCs similar to
the KVs. In the second case we have five independent MCs and the
third case becomes the same as the first case.

\item \qquad {\bf 2}. In this case, we again have three
possibilities. The first case has further two subcases in which
first subcase reduces to (1bi) and the second subcase and the case
(2b) gives 10 independent MCs which contain four KVs and dix
proper MCs. The third case becomes the same as the case (1a).

\item \qquad {\bf 3}. This case trivially gives the same result as
the case (1bi).
\end{description}

It is to be noticed that we have obtained MCs exactly similar to
RCs [14] but with different constraint equations. If these
constraint equations could be solved, then one would be able to
find new exact solutions.

\renewcommand{\theequation}{A\arabic{equation}}
\setcounter{equation}{0}
\section*{Appendix A}

The surviving components of the Ricci tensor are
\begin{eqnarray}
R_{0} & = & {1 \over 4} e^{v - \lambda} (2 v'' + v'^2 - v' \lambda' +
2 \mu' v'), \nonumber \\
R_{1} & = & - {1 \over 4} (2 v'' + v'^2 - \lambda' v' + 4 \mu'' + 2 \mu'^2
- 2 \mu' \lambda'), \nonumber \\
R_{2} & = & - {1 \over 4} e^{\mu - \lambda}
(2 \mu'' + 2 \mu' - \mu' \lambda' + \mu' v') + 1,\nonumber \\
R_{3} & = & R_{2} \sin^2 \theta,
\end{eqnarray}
where prime `$'$' represents derivative w.r.t. $ r $. The
Ricci scalar is given by
\begin{equation}
R = {1 \over 2} e^{-\lambda} (2 v'' + v'^2 - v' \lambda' + 2 \mu' \lambda'
+ 2 \mu' v' + 3 \mu'^2 + 4 \mu'') - 2 e^{-\mu}.
\end{equation}
Using Einstein field equations (1), the non-vanishing components
of energy-momentum tensor $ T_{ab} $ are
\begin{eqnarray}
T_{0} = {e^{v-\lambda} \over 4} (2 \mu' \lambda' - 3 \mu'^2 - 4 \mu'') +
e^{v - \mu},\nonumber \\
T_{1} = {\mu'^2 \over 4} + {\mu' v' \over 2} - e^{\lambda - \mu},\nonumber \\
T_{2} = {e^{\mu-\lambda} \over 4} (2 \mu'' + \mu'^2 - \mu' \lambda' +
\mu' v' + 2 v'' + v'^2 - v' \lambda'),\nonumber \\
T_{3} = T_{2} \sin^2 \theta.
\end{eqnarray}

\renewcommand{\theequation}{B\arabic{equation}}
\setcounter{equation}{0}
\section*{Appendix B}

Linearly independent KVs associated with the spherical symmetry of
the spacetimes are
\begin{eqnarray}
\xi_{(1)} = \sin\phi\partial_\theta+\cot\theta\cos\phi
\partial_\phi,\nonumber\\
\xi_{(2)} = \cos \phi\partial_\theta-\cot\theta\sin\phi
\partial_\phi,\nonumber\\
\xi_{(3)} = \partial_\phi,~\xi_{(4)} = \partial_r
\end{eqnarray}
which are the generators of group $G_4$. The Lie algebra has the
following commutators
\begin{equation}
[\xi_{(1)},\xi_{(2)}] = \xi_{(3)},~~[\xi_{(2)},\xi_{(3)}] =
\xi_{(1)},~~[\xi_{(3)},\xi_{(1)}] =
\xi_{(2)},~~[\xi_{(4)},\xi_{(i)}] =0.
\end{equation}
\newpage

\begin{description}
\item  {\bf Acknowledgments}
\end{description}

The authors would like to thank Higher Education Commission (HEC)
for providing financial assistance during this work. One of us
(MS) is very grateful to Prof. G.S. Hall for his useful comments
during its write up.

\vspace{2cm}

{\bf \large References}

\begin{description}

\item{[1]} Petrov, A.Z.: {\it Einstein Spaces} (Pergamon, Oxford
University Press, 1969).

\item{[2]} Hojman, L. Nunez, L. Patino, A. and Rago, H.: J. Math. Phys.
{\bf 27} (1986)281.

\item{[3]} Green, L.H., Norris, L.K., Oliver, D.R. and
Davis, W.R.: Gen. Rel. Grav. {\bf 8} (1977)731.

\item{[4]} Katzin, G.H., Levine J., and Davis, W.R.: J. Math. Phys.
{\bf 10}(1969)617.

\item{[5]} Misner, C.W., Thorne, K.S. and Wheeler, J.A.: {\it Gravitation}
(W.H. Freeman, San Francisco, 1973).

\item{[6]} Coley, A.A. and Tupper, O.J.: J. Math. Phys. {\bf
30}(1989)2616.

\item{[7]} Carot, J. and da Costa, J.: {\it Procs. of the 6th Canadian
Conf. on General Relativity and Relativistic Astrophysics}, Fields
Inst. Commun. 15, Amer. Math. Soc. WC Providence, RI(1997)179.

\item{[8]} Carot, J., da Costa, J. and Vaz, E.G.L.R.: J. Math. Phys.
{\bf 35}(1994)4832.

\item{[9]} Hall, G.S., Roy, I. and Vaz, L.R.: Gen. Rel and Grav.
{\bf 28}(1996)299.

\item{[10]} Tsamparlis, M., and Apostolopoulos, P.S.:
J. Math. Phys. {\bf 41}(2000)7543.

\item{[11]} Yavuz, \.I., and Camc{\i}, U.: Gen. Rel. Grav.{\bf 28}(1996)691;\\
Camc{\i}, U., Yavuz, \.I., Baysal, H., Tarhan, \.I., and
Y{\i}lmaz, \.I.: Int. J. Mod. Phys. {\bf D10}(2001)751;\\
Camc{\i}, U. and Yavuz, \.I.: Int. J. Mod. Phys.
{\bf D12}(2003)89;\\
Camc{\i}, U. and Barnes, A.: Class. Quant. Grav. {\bf
19}(2002)393.

\item{[12]} Sharif, M.: Nuovo Cimento {\bf B116}(2001)673;\\
Camc{\i}, U. and Sharif, M.: Gen Rel. and Grav. {\bf 35}(2003)97.

\item{[13]} Sharif, M.: Astrophys. Space Sci. {\bf 278}(2001)447.

\item{[14]} Amir, M. Jamil, Bokhari, Ashfaque H. and Qadir,
Asghar: J. Math. Phys. {\bf 35}(1994)3005.

\end{description}

\end{document}